# Intelligent multiscale simulation based on process-guided composite database

Zeliang Liu[1,*], Haoyan Wei[1], Tianyu Huang[1,2], C.T. Wu[1]

*[1] Livermore Software Technology, an ANSYS company, Livermore, CA 94551*
*[2] Northwestern University, Evanston, IL 60208*

---

**Abstract**

   *In the paper, we present an integrated data-driven modeling framework based on process modeling, material homogenization, mechanistic machine learning, and concurrent multiscale simulation. We are interested in the injection-molded short fiber reinforced composites, which have been identified as key material systems in automotive, aerospace, and electronics industries. The molding process induces spatially varying microstructures across various length scales, while the resulting strongly anisotropic and nonlinear material properties are still challenging to be captured by conventional modeling approaches. To prepare the linear elastic training data for our machine learning tasks, Representative Volume Elements (RVE) with different fiber orientations and volume fractions are generated through stochastic reconstruction. More importantly, we utilize the recently proposed Deep Material Network (DMN) to learn the hidden microscale morphologies from data. With essential physics embedded in its building blocks, this data-driven material model can be extrapolated to predict nonlinear material behaviors efficiently and accurately. Through the transfer learning of DMN, we create a unified process-guided material database that covers a full range of geometric descriptors for short fiber reinforced composites. Finally, this unified DMN database is implemented and coupled with macroscale finite element model to enable concurrent multiscale simulations. From our perspective, the proposed framework is also promising in many other emergent multiscale engineering systems, such as additive manufacturing and compressive molding.*

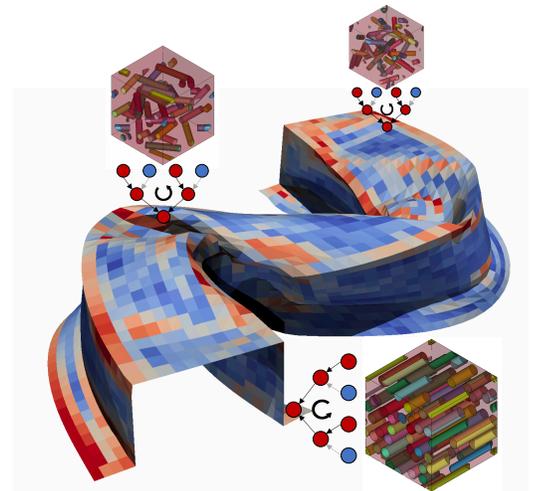

---

## 1. Introduction

   Many current applications in computer aided engineering require an integration of material models at multiple length scales. The complexity comes from the interactions among the fine-scale material phases that vary with the manufacturing process. Due to simplicity and efficiency, micromechanics-based analytical methods (Eshelby, 1957; Mori and Tanaka, 1973; Nemat-Nasser and Hori, 2013) are widely used to approximate the overall material behaviors, but their fundamental assumptions usually regard simplified microstructural morphologies and material models. On the other hand, Representative Volume Element (RVE) with high-fidelity microstructures and nonlinear material behaviors can be analyzed using Direct Numerical Simulations (DNS) tools, such as Finite Element (FE) and Fast Fourier Transformation (FFT)-based methods. In a typical concurrent multiscale simulation, every material point in a macroscale FE model is coupled with a corresponding discretized RVE in the microscale. Depending on the type of the microscale method, it is often referred to as FE$^2$ (Feyel and Chaboche, 2000; Kouznetsova et al., 2002; Feyel, 2003) or FE-FFT (Spahn et al., 2014; Kochmann et al., 2018). However, high CPU time and memory are required to solve these models at different scales simultaneously, which is not affordable if a large-scale model is considered.

---

[*] Corresponding author.
 *E-mail address:* zeliang.liu@ansys.com (Zeliang Liu)

The rise of Machine Learning (ML), especially deep learning based on neural networks, has been continuously advancing the frontier of materials modeling and multiscale simulations. Although the idea of using data-driven models to discover hidden physical relationships and enhance material predictions has been around for decades (Ghaboussi et al., 1991; HKDH, 1999), its popularity was limited mainly due to the lack of computational resources. However, recent evolution of computer hardware systems stimulates the renaissance of many deep learning algorithms (Lecun and Bengio, 1995; Hochreiter and Schmidhuber, 1997). Meanwhile, it allows researchers to explore new ML architectures and algorithms much more effectively (Goodfellow et al., 2014; He et al., 2015), including those designed for materials modeling and multiscale simulations (Liu et al., 2019a; Peng et al. 2020).

## 1.1. Goals of the paper

The main goal of this paper is to present an integrated data-driven modeling framework based on process modeling, material homogenization, mechanistic machine learning, and concurrent multiscale simulation. We consider the simulation to be "intelligent" in the sense that the composite database can make predictions for new microstructures and nonlinear material properties in a non-intrusive way, and the selection of materials models is automatically informed by the manufacturing process. We expect this framework to be applicable to many multiscale engineering systems, such as composite molding, additive manufacturing, and metal forming.

The part of mechanistic machine learning for materials modeling is based upon our recent works on Deep Material Network (DMN) (Liu et al., 2019a; Liu and Wu, 2019). The key features of DMN are the physics-based building block with interpretable fitting parameters, extrapolation capability for material and geometric nonlinearities with only linear elastic training data, and efficient inference with a small number of degrees of freedom. In (Liu et al., 2019b), a transfer-learning approach is proposed to generate a unified set of DMN databases that covers the full-range structure–property relationship. Moreover, DMN enriched by cohesive layers has also been developed for interfacial failure analysis (Liu, 2020). For the first time, we will couple DMN with macroscale FE models for concurrent multiscale simulation.

We will focus on short fiber reinforced composite (SFRC), which we see as an exemplar of multiscale systems with broad industrial applications. For example, the injection-molded SFRC becomes increasingly attractive for lightweight material systems in the automotive industry. However, reliable numerical prediction of SFRC products remains challenging. The nonlinear and anisotropic properties of SFRC are not only affected by the basic properties of each constituent but also strongly influenced by the fiber orientation. Meanwhile, process parameters (e.g., part geometry, position of injection gates, pressure/temperature of the mold, and filling time, etc.) can control the flow of fibers dispersed in the polymer melt, leading to inhomogeneous distribution of fiber orientations in the finished parts (Mortazavian and Fatemi, 2015; Hessman et al., 2019).

The remainder of this paper is organized as follows. In **Section 2**, the proposed data-driven framework to integrate manufacturing process and mechanistic machine learning is discussed. In **Section 3**, we use deep material network to learn a series of short fiber reinforced composites, and we apply the transfer learning approach to derive intermediate material database within the process-guided microstructure space. Examples of concurrent multiscale simulations with highly nonlinear material properties are given in **Section 4**. A brief summary is provided in **Section 5**.

## 1.2 Previous work

Many ML or data-driven methods have been proposed for materials modeling and multiscale simulations. The training data can come from either physical experiments or simulations. One way of mining the data is to perform model reduction on the DNS model of RVEs, such as proper-orthogonal decomposition (Yvonnet and He, 2007; Goury et al., 2016; Fritzen and Kunc, 2018; Rocha et al., 2020), self-consistent clustering analysis (Liu et al., 2016; Liu et al., 2018b; Yu et al., 2019; Gao et al., 2020). Methods have also been formulated based on available stress-strain data – e.g., Gaussian process modeling (Chen et al., 2018; Bostanabad et al., 2018), model-free approaches (Kirchdoerfer and Ortiz, 2016; Ibanez et al., 2018; Eggersmann et al. 2019; He and Chen, 2019), feedforward neural networks (Le et al., 2015; Fritzen et al., 2019; Lu et al., 2019), recurrent neural networks (Wang and Sun, 2018; Ghavamian and Simone, 2019; Frankel et al., 2019), and graph/convolutional neural networks (Frankel et al., 2019; Vlassis et al., 2020). We consider DMN (Liu et al., 2019a; Liu and Wu, 2019; Liu et al., 2019b; Liu, 2020) to be a blend of both approaches: It serves as a reduced-order RVE model while the training only requires stress-strain data.

In terms of injection-molded SFRC, it has been shown that neglecting the process-induced material anisotropies often leads to incorrect prediction of structural failure mode of components (Steinberger et al., 2017), while incorporating every single fiber within a FE model at the structural level will result in prohibitive computational cost. Therefore, upscaling microscopic material information to obtain homogenized material properties via analytical micromechanics approaches (Müller and Böhlke, 2016; Wu et al., 2020) has been quite common for SFRC, where the fiber distributions are characterized by fiber orientation tensors (Advani and Tucker III, 1987; Wang et al., 2018). Along this line, (Adam et al., 2009; Calmels, 2018) imported fiber orientation from injection molding software (e.g., Moldflow, Moldex3D) as an input to the micromechanical software DIGIMAT, which is called by LS-DYNA at each time step to perform material homogenization for the macroscopic stress responses. (Steinberger et al., 2017) calibrated a Mori-Tanaka mean-field homogenization-based anisotropic material model (*MAT_215 of LS-DYNA) against experimental data of the specimen with different fiber orientations. They also performed component-level finite element simulations in LS-DYNA with consideration of fiber distributions. The results demonstrated the effectiveness of the process-integrated multiscale simulation approach.

Nonetheless, improvements to the micromechanics-based material model are needed to reduce deviations from experimental results. While significant progress is made toward multiscale modeling of SFRC products, the highly nonlinear behaviors of SFRC cannot be fully captured through analytical models, which are subject to several assumptions regarding simplified microstructures and material behaviors. Sometimes, the refined analytical models are accommodated specially for SFRC and cannot be easily generalized to other types of material systems. To enhance the predictive capability of multiscale methods, there has been an increasing interest in using DNS to analyze RVE with high-fidelity microstructures. However, DNS tends to be too expensive when used during the entire concurrent multiscale simulation (e.g. $FE^2$, FE-FFT) process, so we propose to employ DNS only during an offline training stage to generate training data for ML purposes.

On the other hand, it is desirable to reduce the number of RVE sampling points. The heterogeneous nature of SFRC structures requires the construction of a huge number of FEA models for RVEs with different microstructures and fiber distributions. Other than computational cost, generation of high-quality three-dimensional meshes for these complex RVEs will be extremely tedious and time-consuming. To this end, (Köbler et al., 2018) proposed to reduce the fiber orientation phase space to a triangle, which is further discretized by a fine triangular mesh. Each nodal point of the triangular sub-cell is linked to a material model with a unique fiber orientation, which is generated by FFT-based computational homogenization during the offline stage. The material behavior in the online structural analysis is estimated by mapping the fiber orientation tensor to the fiber orientation triangle and interpolating the material responses associated with three vertices of each sub-cell. In our work, a similar sampling approach will be adopted.

## 2. The data-driven framework

Predictive multiscale simulation has been a long-standing challenge for various composite material systems due to their hierarchical structures that span multiple length scales. At the same time, the manufacturing process usually introduces spatially varying microstructures with local anisotropy and property difference, which need to be captured in the performance analysis of the final part. An ideal material model should be sensitive to these manufacturing-induced variations, and also accurately predict the overall physical responses in an efficient manner.

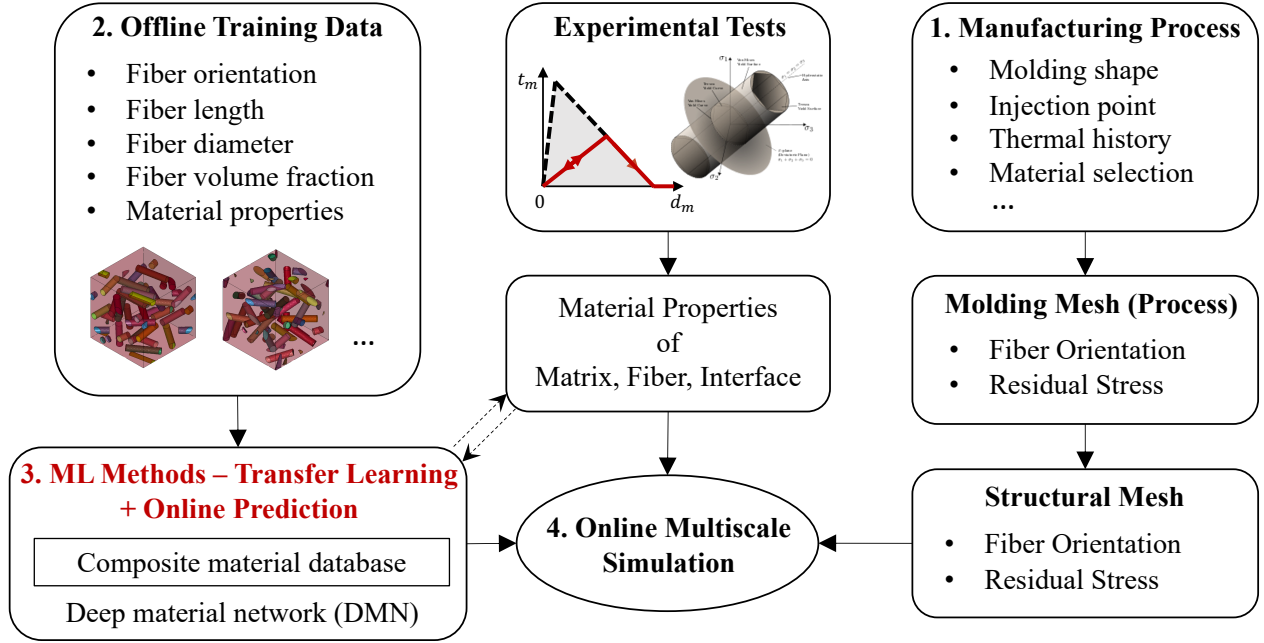

**Fig. 1.** The data-driven framework for multiscale simulation of injection-molded short fiber composites.

We propose a data-driven framework to integrate the manufacturing process and performance analysis based on mechanistic machine learning techniques. In **Fig. 1**, we demonstrate the framework for the injection-molded short fiber composite, and the ultimate goal here is to inform the structural failure analysis with the material microstructure resulting from injection molding simulation. Given the choices of molding shape and different process conditions (e.g., thermal loadings, positions of the injection gates, etc.), a typical molding simulation (**Stage-1**) is able to predict the local fiber orientation and residual stress based on the estimation of interactions between fibers and fluid flow (Wang et al., 2018). After the mapping from the molding mesh to the structural mesh, we can then query the composite material database based on the projected information at each integration point to get the customized material model for performance analysis.

The construction of the composite material database starts from an offline training data generation (**Stage-2**). RVEs are first reconstructed under different geometric descriptors, such as fiber orientation, fiber length, aspect ratio, and phase volume fraction. Samples are randomly generated in a design space spanned by these descriptors, and their ranges are defined either by the theoretical bounds, or from experimental observations. For example, we use the second-order symmetric fiber orientation tensor $\mathbf{A}_{3\times3}$ (Advani and Tucker III, 1987) to represent the distribution of fiber orientation. Through eigen-decomposition, the rotation-free sampling space of the fiber orientation tenser is fully defined by the principal terms $a_{11}$, $a_{22}$, and $a_{33}$, with $a_{11} \geq a_{22} \geq a_{33}$ and $a_{11} + a_{22} + a_{33} = 1$. Meanwhile, the ranges of the fiber length, aspect ratio, and volume fraction can be determined from manufacturing data or experimental microscopy images.

For each RVE model, we build a training dataset by assigning different combinations of material properties to the fiber and matrix phases (Liu et al., 2019a; Liu and Wu, 2019). The purpose here is to explore sufficient cases of phase contrast and material anisotropy for the material network to learn the topological representation. The training of DMN only requires linear elastic RVE data, but it can be extrapolated to predict nonlinear material behaviors in the online stage. As a result, the inputs of each training data point are the stiffness tensors of the fiber and matrix phases, and the output is the homogenized stiffness tensor of the composite through a set of linear elastic RVE analyses under 6 orthogonal loading directions. We use the LS-DYNA RVE package (Liu et al., 2018a) to perform the RVE analyses. Specifically, the "*RVE_ANALYSIS_FEM*" keyword is utilized to define the loading direction and periodic boundary conditions, and the "*DATABASE_RVE*" keyword is used to homogenize the loaded RVE and output the results to a database file automatically. The whole process including RVE reconstruction, meshing, and FE analyses is currently streamlined across different software and programs.

DMN models are trained by gradient-based optimization methods (**Stage-3**) accompanied by transferring the knowledge of pre-trained models to train new RVEs. In addition to accelerating the training process, this so-called "transfer learning" process also guarantees that all trained models of different RVEs share an identical base structure (i.e., the same network topology). In practice, injection molded products contain thousands or even millions of microstructures with different fiber orientations. Obviously, it is not feasible to train an individual model for each microstructure. Therefore, it is necessary to find an effective way to reduce the sampling space (Köbler et al., 2018). To achieve this goal, we propose to utilize transfer learning to generate DMNs for RVEs with intermediate geometry descriptors by simply interpolating only a few pre-trained RVEs, which will be discussed with more details in **Section 3**.

Finding the right phase properties has always been challenging. Experimental tests can be time-consuming and expensive, while some properties (e.g., the interfacial strength) are extremely difficult to measure directly within the material. Our strategy of multiscale material characterization relies on the fact that the material model for a single-phase material is usually available or easy to calibrate. For instance, most material parameters of the pure epoxy matrix and the fiber material, including the elastic constants and the hardening curves, can be measured from experimental tests, and the rest of the parameters can be characterized inversely using available composite coupon test data.

In the last stage (**Stage-4**), we perform the online concurrent multiscale structural analysis driven by the intelligent material models. Although the fitting parameters of the DMN model are optimized via linear elastic data, the essential RVE topology information has been encoded and is still valid in the nonlinear region. The extrapolation capability of DMN makes the proposed data-driven simulation framework non-intrusive in terms of new material behaviors and loading paths that were not considered during the offline training. Since the total number of degrees of freedom in the DMN model is much less than a full-field FEA model associated with the RVE, the computation burden of the concurrent multiscale simulation is much reduced so that solving large-scale practical engineering problems can be achieved. Detailed discussion on the computational performance and cost of multiscale structural simulations will be provided in **Section 4**.

## 3. Deep material networks for short fiber reinforced composites

The DMN training and transfer learning procedures are demonstrated for the short fiber reinforced composite in this section. The same methodology can be also applied to other material systems with manufacturing process-induced morphological variations, such as the nanoparticle reinforced polymer composite and polycrystal materials made by additive manufacturing or metal forming. For demonstration purpose, constant fiber length and diameter are assumed to simplify the sampling space. The RVE geometry, therefore, only depends on the fiber orientation and the fiber volume faction. In addition, the fiber and matrix phases are assumed to be perfectly bonded in the present study, but it is noteworthy to mention that DMN is capable of modeling interfacial behavior through network enrichment (Liu, 2020).

In the online predictions, the orthotropic elastic constants of carbon fibers are assigned to the fiber phase, while the matrix phase is assumed to be an elasto-plastic epoxy material with isotropic hardening. Damage or softening effects are not included in this work, so that the RVE problem is well-posed without the loss of ellipticity. In our future work, non-local or viscous regularization techniques will be introduced for damage and failure analysis.

To minimize the cost of data generation and model training, only 4 RVEs at the vertices of the design space are generated, as shown in **Fig. 2A**. The first three RVEs all have fiber volume fraction $vf = 10\%$. The principal terms of their fiber orientation tensors $[a_{11}, a_{22}, a_{33}]$ are [1/3, 1/3, 1/3], [1/2, 1/2, 0], and [1, 0, 0], which correspond to the 3D random, 2D random, and aligned microstructures, respectively. The fourth RVE is sampled at a higher fiber volume fraction $vf = 30\%$ with aligned unidirectional fibers. Note that the RVE reconstruction and mesh generation processes are not trivial. For example, the fiber packing algorithm may encounter difficulties at high volume fractions, especially for the 3D random and 2D random models.

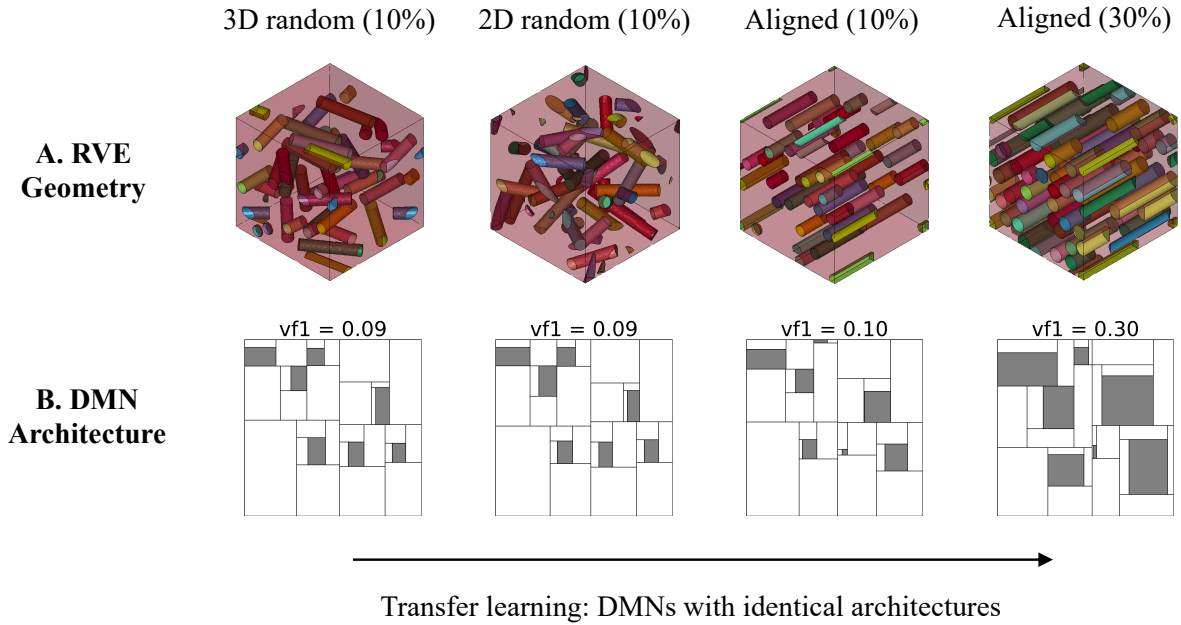

Transfer learning: DMNs with identical architectures

**Fig. 2.** Deep material works for short fiber reinforced composites. **A.** RVE geometries in FE analyses. **B.** Tree-maps of trained DMNs through transfer learning. The volume fraction inferred from the activations $\boldsymbol{z}$ is listed on top of each plot.

For each RVE, we generate 400 training data points and 100 test data points using the full-field FE analysis. Specifically, the test data are added to check the overfitting issue and assess the generalization performance of a trained model. The function to fit in the offline stage is

$$\overline{\mathbf{C}}^{rve} = \mathbf{f}\big(\boldsymbol{z}, \boldsymbol{\alpha}, \boldsymbol{\beta}, \boldsymbol{\gamma};\ \mathbf{C}^{\text{fiber}}, \mathbf{C}^{\text{matrix}}\big),$$

where the $\mathbf{C}^{\text{fiber}}$ and $\mathbf{C}^{\text{matrix}}$ are the input elastic stiffness tensors, and $\overline{\mathbf{C}}^{rve}$ is the output stiffness tensor. The fitting parameters of DMN are $\{\boldsymbol{z}, \boldsymbol{\alpha}, \boldsymbol{\beta}, \boldsymbol{\gamma}\}$, in which the activations $\boldsymbol{z}$ define the weight of each DOF in the network, and $\boldsymbol{\alpha}, \boldsymbol{\beta}, \boldsymbol{\gamma}$ denote the rotation angles at each building block.

Since the fitting parameters are related to the RVE geometry, one can extract useful topological information after training with the mechanical data $\{\mathbf{C}^{\text{fiber}}, \mathbf{C}^{\text{matrix}};\ \overline{\mathbf{C}}^{rve}\}$. In **Fig. 2B**, we plot the tree-maps and estimate the volume fractions of the fiber phase based on the trained activations $\boldsymbol{z}$. The volume fraction inferred from the fitting parameters is very close to the DMN one, even though this information is hidden in the training data. Moreover, we have tried to reproduce the fiber orientation tensor from $\{\boldsymbol{z}, \boldsymbol{\alpha}, \boldsymbol{\beta}, \boldsymbol{\gamma}\}$, and good matches are also observed.

A transfer learning approach is applied here to minimize the training effort for various short fiber reinforced RVEs. We start from the 3D random RVE with 10% fibers and train it with randomly initialized fitting parameters. The trained model is transferred to initialize the networks for 2D random and aligned RVEs with 10% fibers. Afterwards, the trained aligned model is transferred to the one with a higher fiber volume fraction at 30%. As pointed out earlier, construction of 3D or 2D random RVE geometries with 30% fibers is quite challenging, and difficulties in generating high-quality meshes are often encountered. In this regard, the database interpolation scheme introduced later in this section circumvents these issues and offers a feasible way to generate models for these complex microstructures. All the four RVEs are trained for 10000 epochs. Here, one epoch refers to one round of evaluation on every training data point, which are divided randomly into 20 mini-batches for stochastic gradient decent in the present study.

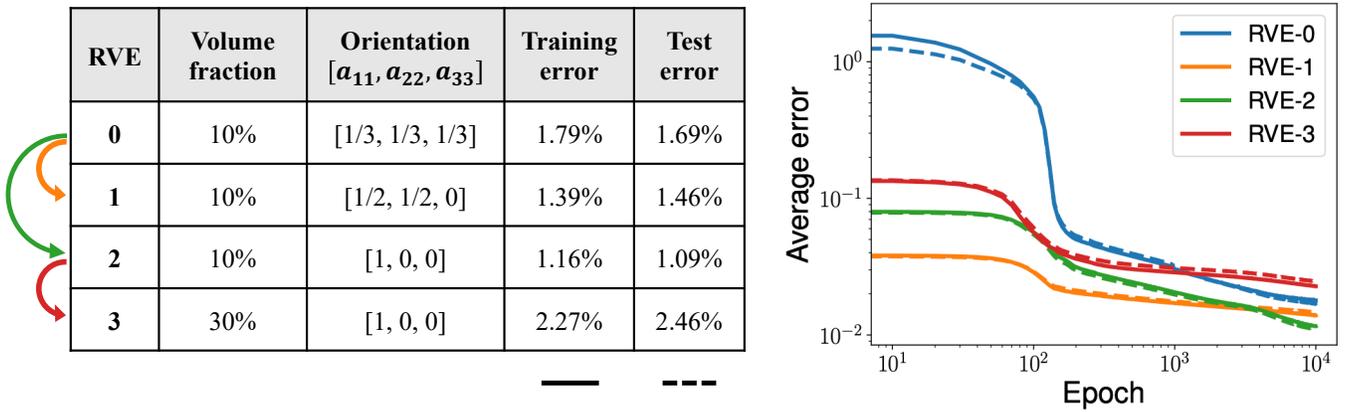

**Fig. 3.** Training history and results for four RVEs on the vertices of the sampling space. $[a_{11}, a_{22}, a_{33}]$ are the principal terms of the fiber orientation tensor. The arrows illustrate the transfer learning path, and RVE-0 is trained with random initialization. The histories of training errors and test errors are denoted by solid lines and dashed lines, respectively.

We set the number of layers in DMN to 8, so that each DMN originally contains 128 DOFs. Since node deactivation and model compression (Liu et al., 2019a; Liu and Wu, 2019) are enabled, the final number of DOFs is reduced to 33 per DMN after training, with 7 for the fiber phase and 26 for the matrix phase, which can significantly save computational storage and enhance efficiency for the online prediction. Note that these numbers can be counted from the tree-map plots in **Fig. 2B**.

The training results and histories are summarized in **Fig. 3**, where the accuracy is measured by the scaled mean absolute error. As we can see from the table, all the 10% RVEs reach a training error less than 1.8%, whereas the responses of the 30% RVE appear more difficult to be captured under similar network complexity. The histories of the average training and testing errors are presented in the curves of **Fig. 3**. The first RVE (i.e., RVE-0) begins with a large training error since the fitting parameters are randomly initialized without any pre-knowledge about the microstructure. For the other three RVEs, the trainings start from much lower errors around 10% because of the knowledge obtained from the pre-trained network via transfer learning.

Another important observation from **Fig. 3** is that test error on unseen data points is almost the same as the training error, indicating that there is no overfitting issue. This nice feature of DMN is attributed to the essential physics embedded in the two-layer building block, which also enables its extrapolation capability to unknown materials and loading spaces.

Since the base tree structures of all the DMNs obtained from transfer learning are analogous, a continuous migration between different database can be realized through direct interpolation of the fitting parameters. For a new intermediate data point $(\mathbf{X}^*, \mathbf{Y}^*)$, where the superscript (*) denotes the intermediate state, the regression function can be written as

$$\mathbf{Y}^*(\mathbf{X}^*) = \mathbf{r}\big(\mathbf{X}^* | (\mathbf{X}_1, \mathbf{Y}_1), (\mathbf{X}_2, \mathbf{Y}_2), \ldots, (\mathbf{X}_N, \mathbf{Y}_N)\big),$$

where the dependent variables $\mathbf{Y}_i$ are the DMN fitting parameters, and the independent variables $\mathbf{X}_i$ are the geometric descriptors of the short fiber reinforced composite:

$$\mathbf{Y}_i = [\mathbf{z}, \boldsymbol{\alpha}, \boldsymbol{\beta}, \boldsymbol{\gamma}]_i \,, \quad \mathbf{X}_i = [vf, a_{11}, a_{22}]_i \,.$$

To determine the unknown parameters for a linear regression model with three independent geometric descriptors (i.e., $[vf, a_{11}, a_{22}]^*$), we need four linearly independent data pairs $(\mathbf{X}_i, \mathbf{Y}_i)$, which correspond to four trained RVEs (i.e., $N = 4$ in the above equation). It is noteworthy that it is possible to extrapolate the database to high volume fraction region outside of the sampling space (e.g., $vf = 50\%$) (Liu et al., 2019b), but constructing a reference full-field DNS model with densely packed fibers is challenging due to both reconstruction and meshing difficulties. Therefore, we limit the fiber volume fraction in the range from 10% to 30% for the present study.

After offline training with elastic data and transfer learning, we use the unified DMN database to predict nonlinear material behaviors. The fiber phase is considered to be orthotropic linear elastic with high strength in the fiber direction. The matrix phase is modeled as an elasto-plastic epoxy material with an exponential hardening law:

$$\sigma^Y(\varepsilon^p) = -a_2 \exp(-a_1 \varepsilon^p) + a_3,$$

where $\sigma^Y$ denotes the yielding stress, and $\varepsilon^p$ denotes the effective plastic strain. All the material constants of the fiber and matrix phases are listed in **Tab. 1**. It can be seen that the matrix starts to yield at $a_3 - a_2 = 30$ MPa, and it becomes close to perfectly plastic under relatively large plastic deformation. The fiber and matrix phases are assumed to be perfectly bonded.

**Tab. 1.** Material parameters of the short fiber reinforced composites in the online prediction stage.

| | $E_1$ (GPa) | $E_2$ (GPa) | $E_3$ (GPa) | $G_{12}$ (GPa) | $G_{13}$ (GPa) |
|---|---|---|---|---|---|
| **Short fiber** | 245.0 | 19.8 | 19.8 | 29.2 | 29.2 |
| | $G_{23}$ (GPa) | $\nu_{12}$ | $\nu_{13}$ | $\nu_{23}$ | |
| | 5.9 | 0.023 | 0.023 | 0.670 | |
| **Matrix** | $E_m$ (GPa) | $\nu_m$ | $a_1$ | $a_2$ (GPa) | $a_3$ (GPa) |
| | 3.8 | 0.387 | 140 | 0.09 | 0.12 |

The loading path in each direction is randomly generated. To obtain plastic-dominant responses, we constrain the stress $\sigma_{33}$ to vanish. This way the influence of hydrostatic loading, which is almost elastic, can be minimized. Note that this constraint is not prescribed in the concurrent simulations to appear in **Section 4**, where each microscale DMN receives all 6 strain components from the macroscale model and returns 6 components of the homogenized stress.

**Fig. 4** presents the comparison between stress-strain curves from DMN and DNS. Two RVE geometries and two loading paths are considered. The first RVE (in the left column) has a 3D random structure with 10% fiber volume fraction, which has been trained explicitly in the offline stage. The DMN model for the second

RVE is obtained from the database interpolation without training. As we can see from the plots, the DMN models predict the nonlinear material response very well. Other than insufficient DMN layers, the slight mismatch may result from the element locking issue in the DNS model under plastic deformations, as the adopted standard linear tetrahedron finite elements can give over-stiff response when the material is nearly incompressible. On the other hand, the DMN models do not suffer from locking issues as element discretization is avoided. The accuracy of offline training can be further improved by adopting hexahedral/high-order tetrahedral elements with reduced integration/pressure projection or other techniques to alleviate locking. Nevertheless, the whole data-driven framework based on DMN should still apply.

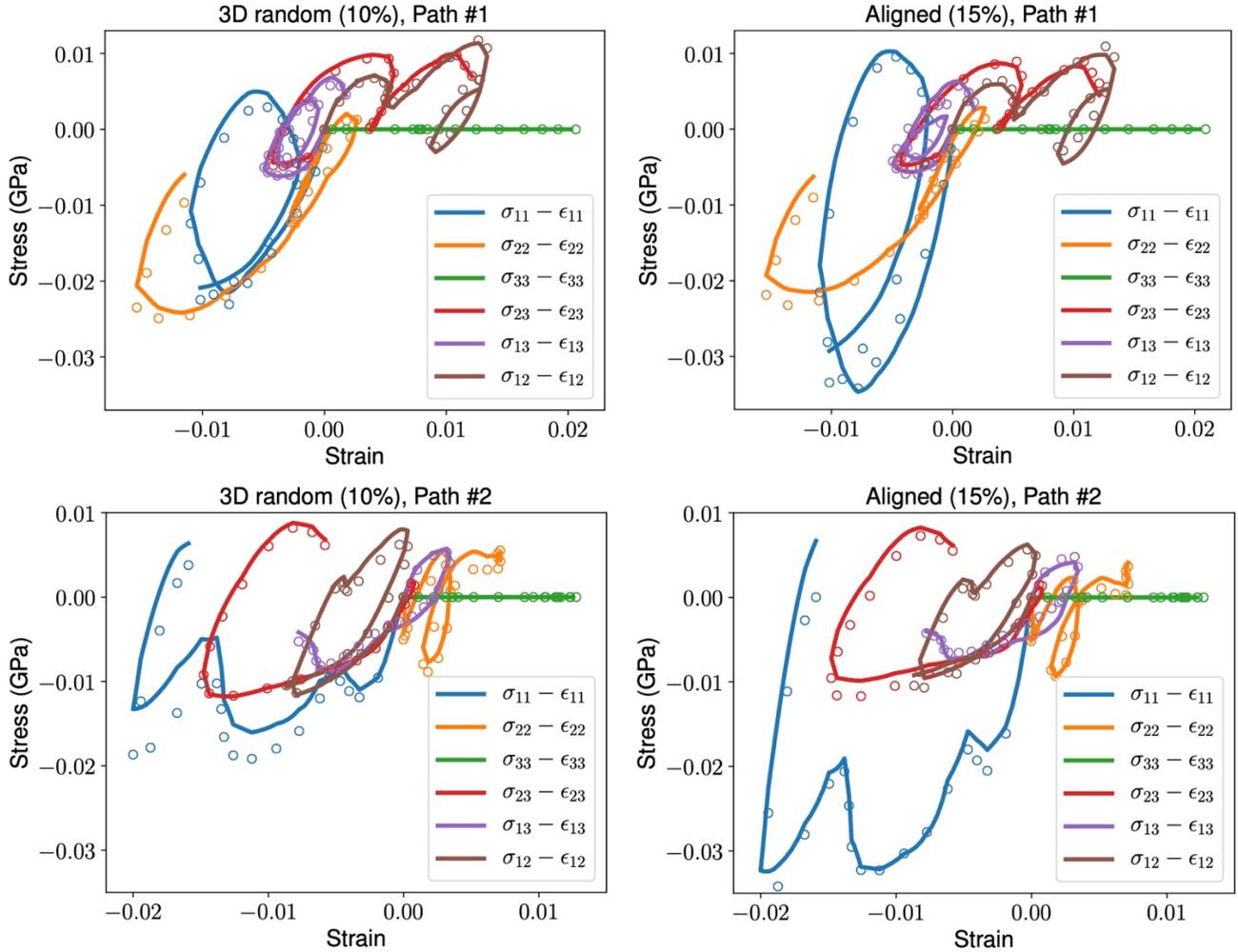

**Fig. 4.** Stress-strain curves predicted by DMN and DNS for two RVE geometries (left-right) under two randomly generated loading paths (top-down). The DMN model with aligned fibers and $vf = 15\%$ (right) is obtained by database interpolation without training. The stress in 33 direction is set to 0. The DMN results are denoted by the solid lines, and the DNS results are denoted by the circles.

In the online prediction stage, DMN is much more efficient than DNS. A typical LS-DYNA FE model for RVE has around 360,000 linear tetrahedron elements, and it takes about 1100 seconds to complete a loading path shown in **Fig. 4** (20 loading steps in an implicit quasi-static analysis) using 8 CPUs. In comparison, the DMN model only has 33 DOFs, and the analysis under the same loading path takes 3 seconds using 1 CPU on the same workstation. The significant reduction in computational cost enabled by DMN opens the possibility of performing concurrent multiscale simulation in a manageable time, and we expect the ratio of cost reduction will be even more significant after migrating the DMN functions from Python to Fortran environment.

## 4. Multiscale concurrent simulation driven by DMN database

The DMN database is implemented and linked to LS-DYNA for concurrent multiscale structural simulations. While the microscale DMN model is always implicit, the macroscale model can be either explicit or implicit, since both stress and tangent stiffness tensor are given by the top node of DMN. To demonstrate this new structural analysis capability, we will focus on multiscale explicit dynamic FEA examples.

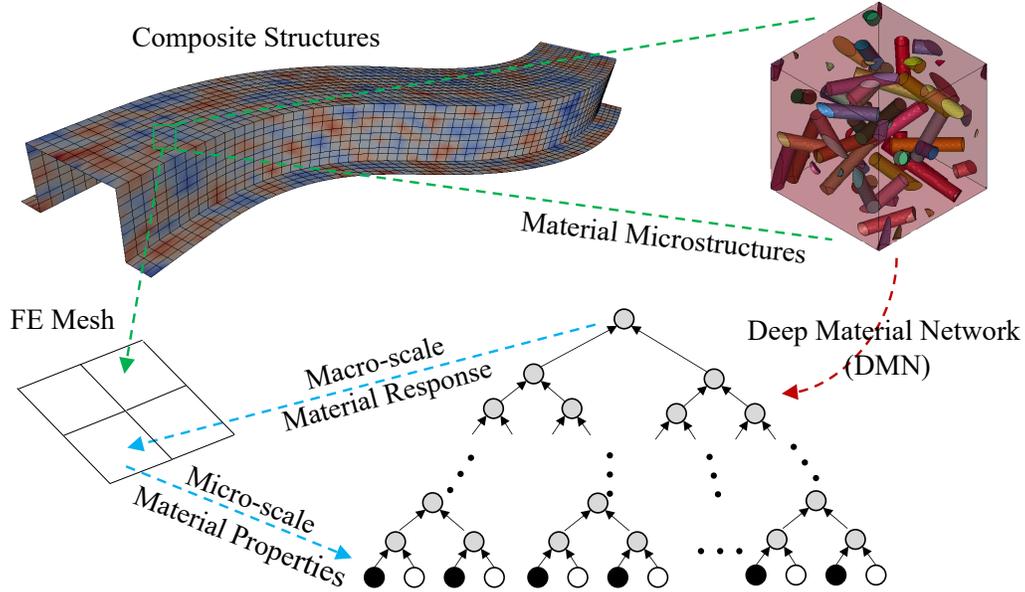

**Fig. 5.** Schematic of the DMN-based concurrent multiscale structural analysis. Each integration point in the macroscale composite structure is linked to a microscale DMN model. The macroscale material responses (e.g., stress) are extracted from the top node of DMN, while the internal variables locate at the nodes in the bottom layer.

As depicted in **Fig. 5**, DMN is applied at each time step during the structural simulation, which involves forward homogenization/rotation and backward de-homogenization/rotation of material information at integration points of finite elements, yielding path-dependent material constitutive response for the overall composite materials, and accordingly, material state variables associated with the bottom layer nodes of the network are dynamically updated and stored. As such, the complex nonlinear material response and structural deformation process are well-captured simultaneously.

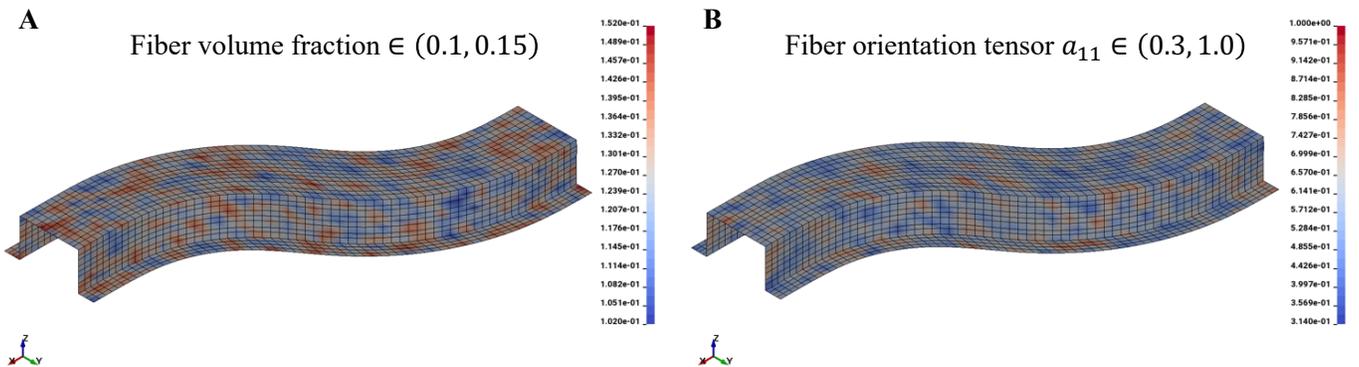

**Fig. 6.** Random distribution of microscale geometric descriptors in the S-rail finite element model. **A.** Fiber volume fraction $vf$. **B.** The $a_{11}$ component in the fiber orientation tensor.

To demonstrate this new structural analysis capability, multiscale explicit dynamic FEA examples are presented as follows. Firstly, we model an S-rail structure shown in **Fig. 6**. Initial velocities along x-direction are prescribed at cross sections of both ends, leading to large deformations and localized plastic strains. The channel is made of short fiber reinforced plastics, for which the actual material microstructural information can be obtained either from image data analysis, or from injection molding simulation packages (such as Moldex3D, MoldFlow, etc.). Herein, for demonstration purpose, we neglect the actual material manufacturing process, and simply assume heterogenous material distribution by assigning random fiber volume fractions and orientations throughout the whole structure, as illustrated in **Fig. 6**.

The channel is discretized with 2340 Belytschko-Tsai shell elements. Specifically, the type 25 shell is chosen to directly adopt fully 3D constitutive responses of DMN without ad-hoc modifications. Generalized plane stress conditions in DMN can also be imposed to yield constitutive response compatible with conventional thin shell elements (e.g., Type 2), but the results are not reported here. Efficient one-point quadrature scheme is adopted, where two through-thickness integration points are used per element, although more integration points can be chosen, as DMN does not add any restrictions on quadrature rules.

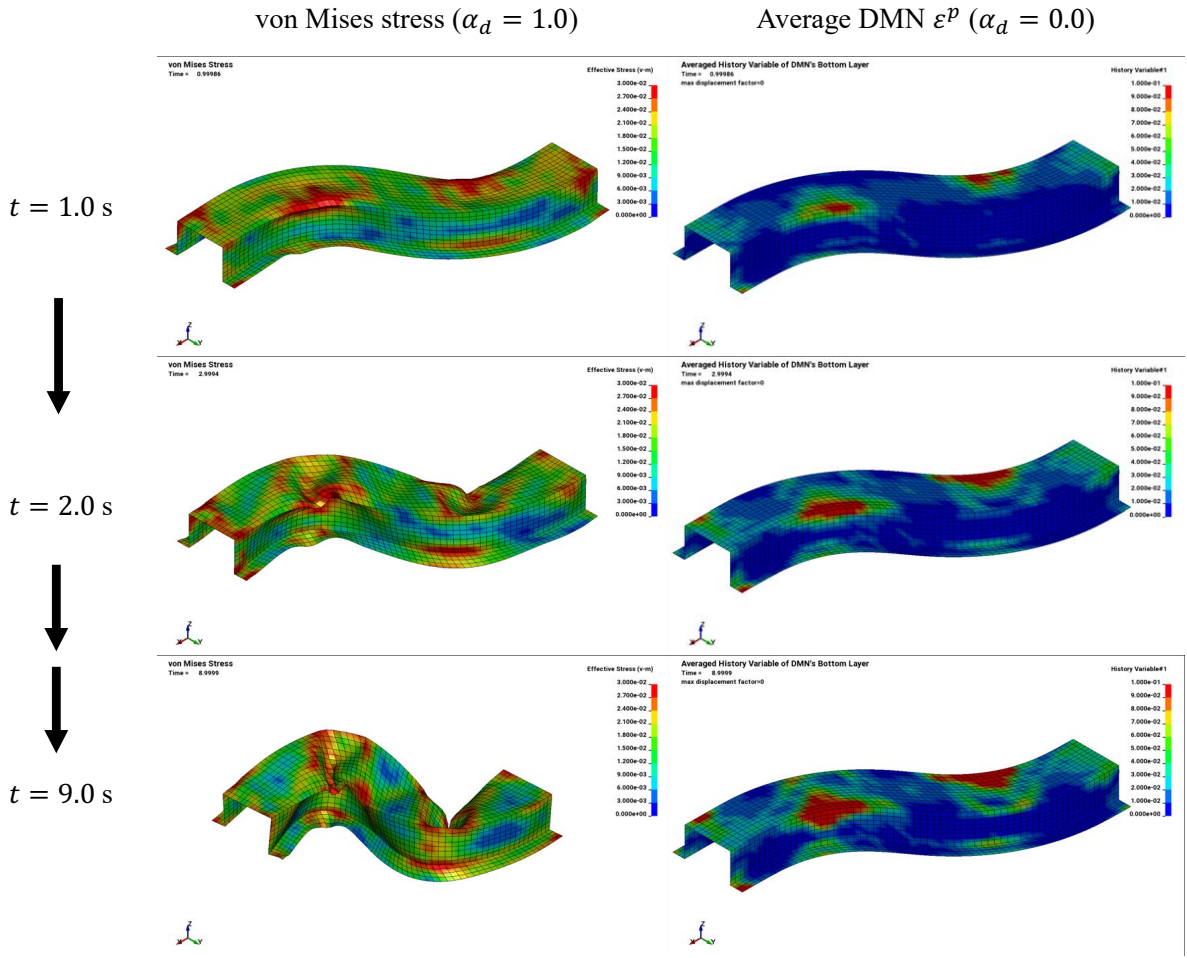

**Fig. 7.** Results of the S-rail structure. **A.** von Mises stress (unit: GPa) distribution at the upper surface of shell elements during the S-rail deformation process (displacement scale factor: $\alpha_d = 1.0$). **B.** Evolution of averaged material state variables of bottom layer nodes of DMN at the upper surface of shell elements (displacement scale factor: $\alpha_d = 0.0$).

Based on the local material microstructural information, a DMN database specific to every integration point is generated through transfer learning of the pre-trained networks and database interpolation as discussed in **Section 3**. It is noteworthy to mention that, while simple linear elastic constitutive laws are adopted during the offline stage to learn the essential features and physics of composite materials, we employ an orthotropic elastic model for fibers and an isotropic hardening plasticity model for the epoxy at the online structural analysis stage to capture highly nonlinear material behaviors.

**Fig. 7** shows the predicted large deformations of the S-rail at several time instances, and in **Fig. 7A**, the distributions of the von Mises stress are plotted. **Fig. 7B** shows the weighted effective plastic strain $\varepsilon^p$ at the bottom layer of each DMN. Comparing with the von Mises stress distribution, we can clearly see that path-dependent localized plastic deformations of the short fiber reinforced composite structures are effectively captured by the physics-embedded machine learning algorithm.

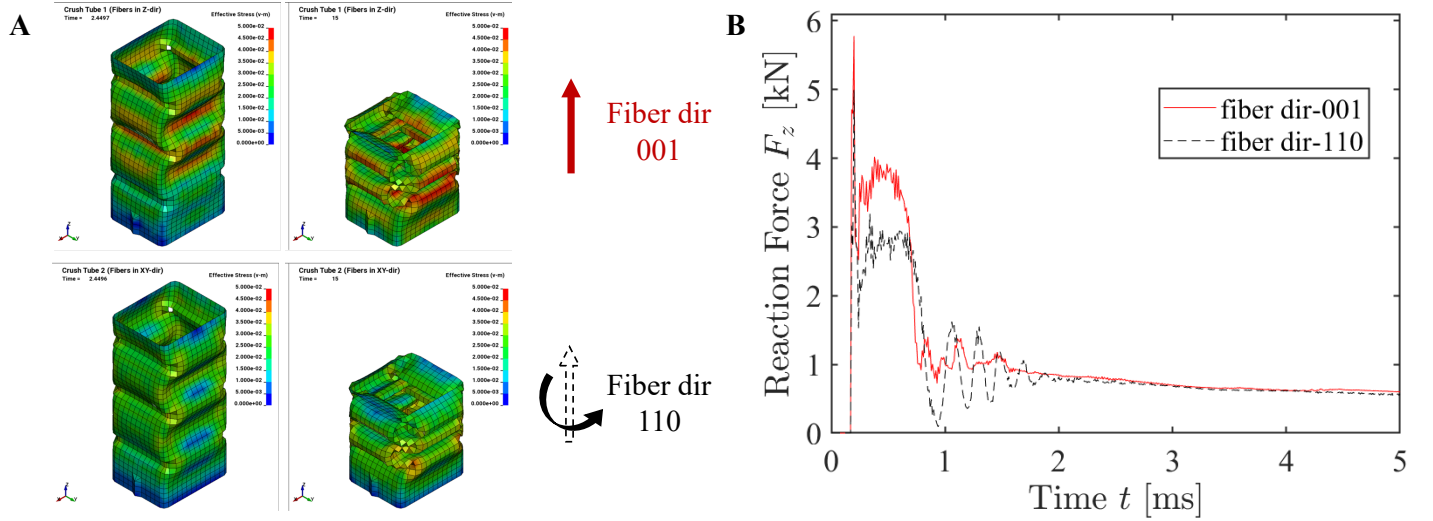

**Fig. 8.** Results of the crush tubes. **A.** von Mises stress (unit: GPa) distribution at the upper surface of shell elements during tube crash processes (top row: fibers along the vertical Z-direction; bottom row: fibers in the horizontal X-Y plane, displacement scale factor: $\alpha_d = 1.0$). **B.** Force-displacement curves.

Next, we performed concurrent multiscale simulations of fiber-reinforced composite tubes under compressive crash loadings, which we refer to as "crush tube" later in the paper. Two crush tubes with uniform material distributions are considered. They have the same fiber volume fraction $vf = 15.2\%$ but different fiber orientations as shown in **Fig. 8A**. The tube is discretized with 641 Belytschko-Tsai shell elements, with designed imperfections to control the buckling mode. A rigid wall with an initial velocity presses the crush tube from the top through surface contact. The reaction force and displacement can be measured on the rigid wall in the simulation.

As shown in the contour plots, the tube with short fibers along the vertical Z-direction yield higher stress distributions than the tube with fibers in the horizontal X-Y plane, which is expected as the simulation is under a vertical compressive motion control. This difference due to multiscale microstructural effect is also observed in the force-displacement curves given in **Fig. 8B**. The tube with fiber along the loading direction experiences a higher post-buckling force. The results are promising as the microstructure-sensitive stress-strain responses are purely given by the DMN database; thus, no empirical material model is needed in the macroscale FE simulation. We are currently integrating the multiscale simulation framework with injection molding simulation results to take manufacturing effects into account for the structural failure analysis, and we will utilize experimental data to further calibrate and validate the developed multiscale modeling approach.

**Tab. 2.** Computational times of the concurrent multiscale simulations. The microscale DMNs have 8 layers and 33 DOFs. All the computations are performed on a workstation with 20 Intel® Xeon® CPU E5-2640 v4 2.40 GHz processors.

| Macroscale model | | | Computational wall time (min) | | |
|---|---|---|---|---|---|
| Name | Num of Elements | Loading cycles | $N_{CPU} = 1$ | $N_{CPU} = 4$ | $N_{CPU} = 16$ |
| S-rail | 2340 | 17372 | 224.0 | 65.5 | 18.1 |
| Crush tube | 641 (Sym) | 17640 | 68.4 | 24.5 | 11.0 |

The computational times for the concurrent simulations of the aforementioned examples are summarized in **Tab. 2**. Since most computational resources are consumed by the DMN model evaluation at every macroscale element, the wall time scales well with the number of CPUs ($N_{CPU}$). We expect the wall time will scale closer to linearly for larger models. Let us make a naïve estimation based on the wall time of the S-rail model on 4 CPUs here: If a structure-level model has **1,000,000 finite elements** and the same number of loading cycles, the simulation would take around **19.5 hours to finish using 96 CPUs**. Meanwhile, code optimization and parallelization (e.g. GPU) are being performed to further accelerate the computation.

## 5. Conclusion

In this paper, we have shown that the concurrent multiscale simulations are perfectly approachable using computational homogenization and mechanistic machine learning for short-fiber reinforced composites. We consider the simulations to be "intelligent" for two reasons: 1) The material model for any microstructure induced by the manufacturing process can be extracted from the composite database; 2) The Deep Material Network (DMN) can capture nonlinear material behaviors under complex loading paths by learning the essential topological representation only from linear elastic data.

Training processes of DMN show good convergence for various short fiber reinforced RVEs with different fiber orientations and volume fractions. To reduce the number of sampling points, we employed the transfer learning approach and generated DMNs for intermediate RVE geometries by simply interpolating a few pre-trained RVEs. Moreover, DMN shines in the case of extrapolation to nonlinear RVEs with elasto-plastic matrix under randomly generated loading paths. It is shown to achieve good accuracy against DNS results using much less degrees of freedom. This is especially important for efficient concurrent multiscale simulations as we demonstrated for two structural models with explicit dynamics.

We believe this work will push machine learning and data-driven materials modeling to computer-aided engineering at an industrial scale. At the same time, we will integrate the multiscale simulation with process simulation software and validate the methods using experimental testing data. Further enhancements of our models are also promising. For example, in (Liu, 2020), interfacial behaviors (e.g., debonding) between the fiber and matrix phases can be considered. Moreover, implicit macroscale analysis is ready to be implemented, which will be advantageous for multiscale simulation of longer-time events.

It should be noted that the proposed data-driven framework based on DMN is quite general. It can be applied to different material systems in addition to the short fiber reinforced composite, such as the nanoparticle reinforced rubber composite, polycrystalline materials, and different types of carbon fiber reinforced composites (Liu and Wu, 2019). In terms of the manufacturing processes, high-throughput microstructure prediction and characterization are undergoing intensive research (Kaufmann et al., 2020), and we are currently working along this direction.

## Acknowledgements

The authors would like to thank Dr. Nielen Stander and Dr. Anirban Basudhar for helpful discussions. The authors warmly thank Dr. John O. Hallquist of LST for his support to this work. A part of the work will be presented at the 16[th] LS-DYNA user conference.